# Collecting large-scale publication data at the level of individual researchers: A practical proposal for author name disambiguation[1]

Ciriaco Andrea D'Angelo

*Dept of Engineering and Management*
*University of Rome "Tor Vergata", Rome, Italy*
dangelo@dii.uniroma2.it

Nees Jan van Eck

*Centre for Science and Technology Studies*
*Leiden University, Leiden, The Netherlands*
ecknjpvan@cwts.leidenuniv.nl

**Abstract**

The disambiguation of author names is an important and challenging task in bibliometrics. We propose an approach that relies on an external source of information for selecting and validating clusters of publications identified through an unsupervised author name disambiguation method. The application of the proposed approach to a random sample of Italian scholars shows encouraging results, with an overall precision, recall, and F-Measure of over 96%. The proposed approach can serve as a starting point for large-scale census of publication portfolios for bibliometric analyses at the level of individual researchers.

# 1. Introduction

One of the first steps in bibliometric evaluation involves collecting the census of publications produced by the subjects included in the evaluation. This census must obviously be complete in terms of representing the true publication portfolio of the subjects in question, whether they be individual researchers, research groups, organizations, territories, or nations. The outcomes of a bibliometric research evaluation (especially if carried out at the individual level) are reliable only if based on high-quality datasets, which typically are difficult to extract from the main bibliometric data sources (Schulz, 2016). Depending on the bibliometric data source used, the problem of identifying all the publications produced by a person or unit of interest is more or less complex and never trivial.

The disambiguation of the true identity of an author of a publication extracted from a bibliometric data source is in fact a process with many pitfalls because of the following reasons:
- Lack of standardization in identifying the authors' institutional affiliations (Huang, Yang, Yan, & Rousseau, 2014; Morillo, Santabárbara, & Aparicio, 2013);
- Variability in naming a single person in different publication bylines (Cornell, 1982);
- Errors in transcribing names; and
- Problems of homonymy which, in certain contexts, can be extremely frequent and very difficult to solve (Aksnes, 2008).

The most frequently used indicators to measure the reliability of bibliometric datasets are precision and recall, which originate from the field of information retrieval (Hjørland, 2010). Precision is the fraction of retrieved instances that are relevant while recall is the fraction of relevant instances that are retrieved. Their values depend on the presence of two types of errors:
- "False positives" or publications assigned to a given subject while the subject has in fact not authored them; and
- "False negatives" or publications not assigned to the evaluated subject while the subject in fact has authored them.

The evaluator's aim is to construct a bibliometric dataset in which both types of errors can be reduced to acceptable levels. For this purpose, in a large-scale bibliometric evaluation, the evaluators have at least three different options:
1. They can ask the subjects being evaluated to submit their publications;
2. They may first draw a list of unique author identifiers and then use this information to query a bibliometric database; or
3. They can extract publications in the period of interest from a bibliometric database and, then, disambiguate the true identity of the relevant authors.

These approaches present significant trade-offs both in terms of precision/recall and cost.

*Publication lists prepared and submitted by the assessed entity*

This type of approach can guarantee a high level of precision and recall since, at least in theory, no one is more qualified than the subjects themselves to produce a publication list that can meet the specifications provided by the evaluator. However, this is a particularly "costly" approach because of the opportunity cost of research foregone by the surveyed subjects for collecting and selecting outputs for the evaluation. Savings can be achieved by avoiding the direct involvement of subjects to be evaluated, however, any type of savings would then have to be balanced against the reduction in precision and recall for the final dataset (Hicks, 2009; Harman, 2000).

*Relying on unique author identifiers*
  The introduction of unique identifiers for scientific entities (researchers, publications, organizations, etc.) is important and necessary for improving the quality of information systems (Enserink, 2009). For individual scientists, the challenge is very complex and the stakes high, which can be witnessed by the rapid progress of attempts for global identification of scientists (Mazov & Gureev, 2014). The global bibliometric databases, Scopus by Elsevier and Web of Science (WoS) by Clarivate Analytics, provide functions for authors to register their publications. The registry of Scopus consists of the so-called Scopus Author Identifiers while the registry of WoS of ResearcherIDs. ORCID (Open Researcher and Contributor ID) is another registry that needs to be mentioned. ORCID aims to "…create a world in which all who participate in research, scholarship and innovation are uniquely identified and connected to their contributions and affiliations, across disciplines, borders, and time" (Haak, Fenner, Paglione, Pentz, & Ratner, 2012). For such registries to work, most authors would have to participate. At the moment, this is not the case, since the penetration is often insufficient and not uniform in terms of the country and/or field (Youtie, Carley, Porter, & Shapira, 2017).

*Setting up a large-scale bibliometric database in desk mode*
  The evaluator could proceed by autonomously collecting publications produced by the subjects from relevant bibliometric databases. They would have to query the database, limit the results by the publication window of interest and the country of the authors who need to be analyzed, and successively disambiguate the true identity of the authors of the extracted publications for the identification of the subjects of interest.
  This option offers rapid and economical implementation, not requiring the support of the evaluated subjects, as for the first two approaches. However, as said, the census of the scientific outputs of single identifiable individuals is challenging because of homonyms in author names and variations in the way authors indicate their name and affiliation (Smalheiser & Torvik, 2009). Methods to disambiguate author names are usually categorized as supervised or unsupervised. Supervised methods require manually labeled data to train an algorithm. The need for training data makes this approach expensive in practice In fact, the manual labeling of data rapidly becomes impractical for large-scale bibliometric databases and maintaining the training data can be prohibitive when the data changes frequently. Unsupervised approaches do not need manually labeled data. Instead, they formulate the author-name disambiguation problem as a clustering task, where each cluster contains all publications written by a specific author. Important shortcomings in existing unsupervised approaches include poor scalability and expandability. To address such challenges, Caron and Van Eck (2014) proposed a rule-based scoring and oeuvre identification method (from now on the CvE method) to disambiguate authors in the in-house WoS database of the Centre for Science and Technology Studies (CWTS) at Leiden University. The results of this method have been used in several studies, including studies on contributorship, collaboration, research productivity, and scientific mobility (e.g., Chinchilla-Rodríguez, Bu, Robinson-García, Costas, & Sugimoto, 2018; Chinchilla-Rodríguez at al., 2018; Larivière & Costas, 2016; Larivière et al., 2016; Palmblad & Van Eck, 2018; Robinson-Garcia et al., 2019; Ruiz-Castillo & Costas, 2014; Sugimoto et al., 2017; Tijssen & Yegros, 2017). In a recent study (Tekles & Bornmann, 2019), the approach by CvE was compared with several other unsupervised author name disambiguation approaches based on a large validation set containing more than one million author mentions. It turned out that the CvE approach outperforms all other approaches included in the study.

Both supervised and unsupervised approaches generally tend to favor precision over recall. In fact, in the CvE approach, the publication oeuvre of an author can be split over multiple clusters of publications if not enough proof is found for joining publications together. This means that the results of the method are not immediately usable for evaluative purposes, unless a further step of re-aggregation of the split publication oeuvres is carried out. This step can be carried out only using some external source of information. D'Angelo, Giuffrida, and Abramo (2011) proposed a method that links a bibliometric database to a reference institutional database providing information on the university affiliation and research field of each Italian academic professor in order to disambiguate their authorship in the WoS (from now on the DGA method).

Starting from the authors' experience, in this paper we propose a new approach in which the author name disambiguation results of the CvE method are filtered and merged based on information retrieved from a reference institutional database originally used in the DGA method. Different from most contributions dedicated to author name disambiguation in the literature, we will apply our approach not to a "standard" dataset already used for validation purpose by other scholars. To demonstrate the potential value of the proposed approach in real research evaluation exercises, it will be applied to a dataset containing 615 randomly selected Italian academic scholars. More specifically:
- Personal information on the scholars retrieved from the external database will be used to extract and validate the publication oeuvres identified using the CvE method;
- The precision and recall of three different "filtering" scenarios will be measured; and
- The results obtained in the three scenarios will be compared with three distinct baselines. The DGA method will be used as one of the baselines.

Even though it is based on a limited randomly extracted sample, this work can be useful for anyone carrying out a large-scale census of scientific publications (research managers, policy makers, and evaluators in general struggling with performance assessment at the individual level) by providing empirical measures of accuracy of different usage options of the CvE method. Of course, some additional data at the individual level has to be available, however, as we will demonstrate, these are simple lists containing, for each researcher some basic data, i.e. the name and their affiliation city.

The rest of this paper is organized as follows. Section 2 presents a summary of the state of the art in author name disambiguation approaches in bibliometrics. Section 3 describes the method and dataset used in our study. Section 4 presents the results obtained by comparing different validation criteria of publication oeuvres retrieved for each of the subjects in the dataset. The closing section provides some final remarks.

## 2. Approaches to author name disambiguation

The disambiguation of author names has been recognized as an important and challenging task in the field of bibliometrics, digital libraries, and beyond. When bibliometric studies include many researchers, it is unfeasible to perform disambiguation manually. Automatic approaches to disambiguate author names have therefore been developed. Many different approaches have been proposed in the literature (Ferreira, Gonçalves, & Laender, 2012). What is common between all the approaches is that they use some measure of similarity to identify publications most-likely authored by the same individual. One way to distinguish approaches from each other is to categorize them as supervised or unsupervised (Smalheiser & Torvik, 2009). In this section, we briefly discuss these different type of approaches. We refer the reader to works of Cota, Ferreira, Gonçalves, Laender, and Nascimento (2010); Ferreira, Gonçalves, and Laender (2012); and Smalheiser and Torvik (2009) for a more detailed discussion.

Supervised approaches use pre-labeled training data to train the parameters of a machine learning model to either predict the author of a publication (e.g. Ferreira, Veloso, Gonçalves, & Laender, 2010; Han, Giles, Zha, Li, & Tsioutsiouliklis, 2004; Veloso, Ferreira, Gonçalves, Laender, & Meira Jr, 2012) or to determine if two publications are authored by the same individual (e.g. Culotta, Kanani, Hall, Wick, & McCallum, 2007; Huang, Ertekin, & Giles 2006; Smalheiser & Torvik, 2009; Treeratpituk & Giles, 2009). The idea is that after training, the model can be used to disambiguate the authors of sets of unseen publications. Supervised approaches mainly differ in the employed machine learning model (e.g., the Naive Bayes probability model, random forests, or support vector machines) and the publication attributes (e.g., co-authors, affiliations, publication venue, title, keywords, cited references, etc.) considered. The pre-labelled training data is usually a set of publications in which author names have been annotated using unique author identifiers. Although some author name disambiguation datasets are available (e.g., Kim, 2018; Müller, Reitz, & Roy, 2017), getting accurate and unbiased training data is still an important bottleneck in the development of supervised approaches (Song, Kim, & Kim, 2015). For a detailed literature review on this matter, see Kim, Kim, and Owen-Smith, 2019.

In contrast, unsupervised approaches are based on unsupervised techniques such as similarity estimation and clustering (e.g., Cota et al. 2010; Han, Zha, & Giles, 2005; Liu at al., 2014; Schulz, Mazloumian, Petersen, Penner, & Helbing, 2014; Soler 2007; Song, Huang, Councill, Li, & Giles, 2007). A major advantage of unsupervised approaches is that they do not require any pre-labeled training data. Unsupervised approaches typically rely on the similarities between publications to group publications that most likely belong to the same author. Predefined similarity measures (not learned from a training set) consider different information elements (e.g., co-authors, affiliations, publication venue, article title, keywords, cited references, etc.) for calculating the similarity between publications. Unsupervised approaches mainly differ in the way in which the similarity between publications is measured and the used clustering method. Most approaches use agglomerative clustering algorithms such as single-linkage or average-linkage clustering. Similarity measurements vary in the publication attributes that are included, how the attributes are combined, and whether fixed or name dependent similarity threshold values are used to determine if there is enough evidence to assign publications to the same cluster or individual. Name-dependent similarity threshold values can be used to reduce the problem of wrongly merging publication oeuvres of individuals with common names (e.g., Backes, 2018; Caron & Van Eck, 2014).

As seen, both supervised and unsupervised approaches typically rely on the use of various types of publication metadata in addition to the author name itself (Levin,

Krawczyk, Bethard, & Jurafsky, 2012). This includes the names of co-authors, affiliation information, year of publication, publication venue, subject classification, topic as inferred by title, keywords or abstract, and citations to other publications. Author name disambiguation approaches have been applied to the data from various smaller and larger bibliographic databases, including AMiner, CiteSeer, DBLP, PubMed, Scopus, and WoS. It should be noted that not all bibliographic databases contain the same metadata attributes for indexed publications. Missing metadata attributes may impose serious limitations on the accuracy of disambiguation approaches. For instance, if affiliation data or cited reference data is not available in a particular bibliographic database, then this type of information or evidence cannot be exploited to disambiguate authors. In addition to the information stored in bibliographic databases, several studies have explored the possibility to take advantage of external information sources, such as institutional databases (Kawashima & Tomizawa, 2015; D'Angelo, Giuffrida, & Abramo, 2011), the Web (e.g., Abdulhayoglu & Thijs, 2017; Kanani, McCallum, & Pal, 2007; Kang et al., 2009; Pereira et al., 2009; Yang, Peng, Jiang, Lee, & Ho, 2008), or crowdsourcing (Sun, Kaur, Possamai, & Menczer, 2013).

In the following subsections, we describe in more detail the CvE method, the pillar of the proposed approach, and the DGA method, since it is used as one of the baseline methods for evaluating the performance of the proposed approach.

## 2.1 The CvE author name disambiguation method

Figure 1 provides a visual overview of the author disambiguation process followed by CvE (Caron and Van Eck, 2014). Bibliometric metadata related to authors and their publications is taken as input and clusters of publications most likely to be written by the same author are given as output. The CvE method consists of three phases: (1) pre-processing, (2) rule-based scoring and oeuvre identification, and (3) post-processing. The method has been developed to disambiguate all authors in the in-house version of the WoS database available at CWTS. In this paper, the April 2017 version of this database is used. This version of the database includes over 50 million publications indexed in the Science Citation Index Expanded, the Social Sciences Citation Index, and the Arts & Humanities Citation Index.

We now discuss the three phases of the CvE method in more detail. The output of the CvE method consists of an assignment of each publication-author combination in the WoS database to an author oeuvre.

*Figure 1: The CvE author name disambiguation process*

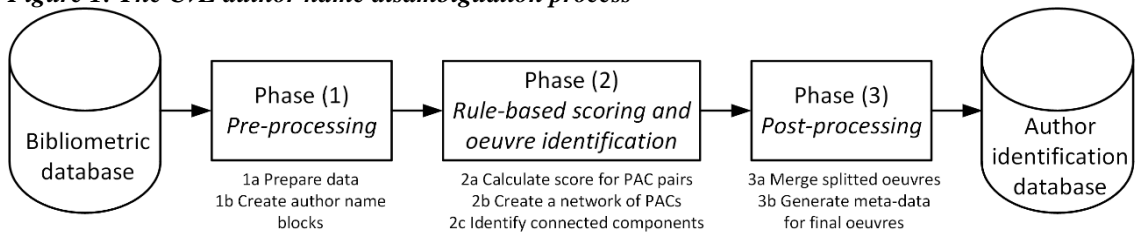

*Pre-processing phase*

In the pre-processing phase, author name blocks are created (On, Lee, Kang, & Mitra, 2005). First, non-alphabetic characters are removed from the names of authors. Next, all author names consisting of the same last name and first initial are assigned to the same author name block. For instance, the author names "Grosso, Andrea Cesare", "Grosso, Andrea", and "Grosso, Anna" are all assigned to the author name block "Grosso, A". The pre-processing phase is important because it leads to a major reduction in computational cost in the next phase.

*Rule-based scoring and oeuvre identification phase*

In the rule-based scoring and oeuvre identification phase, candidate author oeuvres are identified. For each author name block, the corresponding publication-author combinations (PACs) are identified. Next, for each pair of two PACs belonging to the same author name block, a score is calculated. The higher this score, the stronger the evidence that the two PACs belong to the same author oeuvre. If the score of a pair of PACs exceeds a certain threshold, this is considered strong direct evidence that the PACs belong to the same author oeuvre. In this way, a network of PACs is obtained in which two PACs are connected if their score exceeds the threshold. The connected components of this network are identified using single-linkage clustering. The PACs in each connected component are the candidate author oeuvres identified in the rule-based scoring and oeuvre identification phase. Hence, two PACs are assigned to the same candidate author oeuvre if there exists strong direct or indirect evidence to justify this assignment. For instance, suppose there is strong direct evidence that PACs 1 and 2 belong to the same author oeuvre, that PACs 2 and 3 belong to the same author oeuvre, and that PACs 3 and 4 belong to the same author oeuvre. Indirectly, this is then considered strong evidence that PACs 1, 2, 3, and 4 all belong to the same author oeuvre.

The score of a pair of PACs is calculated using a set of scoring rules. The following four types of scoring rules are used:

- *Scoring rules based on comparing author data*. The more similar two authors, the higher the score. The similarity between authors is determined based on their e-mail addresses, their initials, their first names, and their affiliations.
- *Scoring rules based on comparing publication data*. The more similar two publications, the higher the score. The similarity between publications is determined based on shared author names, shared grant numbers, and shared affiliations.
- *Scoring rules based on comparing source data*. The more similar the sources (i.e., journals or book series) in which two publications have appeared, the higher the score. The similarity between sources is determined based on their titles and their WoS subject categories.
- *Scoring rules based on citation relations*. The stronger the citation relatedness of two publications, the higher the score. The citation relatedness of publications is determined based on direct citation links, bibliographic coupling links, and co-citation links.

The score of a pair of PACs is the sum of the scores obtained from the different scoring rules. The scores assigned by each of the scoring rules have been determined based on expert knowledge and have been fine-tuned by evaluating the accuracy of the scoring rules using a test data set. Table 1 presents a detailed overview of all the scoring rules and associated scores. In the case of hyper-authorship and hyper-instituteship publications, the scores of the scoring rules based on shared authors, shared affiliations, and self-citations are lowered. A publication is seen as a hyper-authorship publication if there are at least 50 authors. A publication is seen as a hyper-instituteship publication if there are at least 20 institutes. The lowered scores in the case of hyper-authorship and hyper-instituteship publications are indicated within parentheses in Table 1.

*Table 1: Scoring rules and associated scores in the CvE method*

| Category | Scoring rule | Field | Criterion | Score |
|---|---|---|---|---|
| Author data | 1 | Email | | 100 |
| | 2a | Initials (more than one) | Two initials | 5 |
| | 2b | | More than two initials | 10 |
| | 2c | | Conflicting initials | -10 |
| | 3a | First name | General name | 3 |
| | 3b | | Non-general name | 6 |
| | 4a | Affiliation address (linked to author) | Country, city | 4 |
| | 4b | | Country, city, organization | 7 |
| | 4c | | Country, city, organization, department | 10 |
| Publication data | 5a | Shared co-authors | One | 4 (2) |
| | 5b | | Two | 7 (4) |
| | 5c | | More than two | 10 (5) |
| | 6 | Grant number | | 10 |
| | 7a | Affiliation address (not linked to author) | Country, city | 2 (1) |
| | 7b | | Country, city, organization | 5 (3) |
| | 7c | | Country, city, organization, department | 8 (4) |
| Source data | 8a | Subject category | | 3 |
| | 8b | Journal | | 6 |
| Citation data | 9 | Self-citation | | 10 (5) |
| | 10a | Bibliographic coupling | One | 2 |
| | 10b | | Two | 4 |
| | 10c | | Three | 6 |
| | 10d | | Four | 8 |
| | 10e | | More than four | 10 |
| | 11a | Co-citation | One | 2 |
| | 11b | | Two | 3 |
| | 11c | | Three | 4 |
| | 11d | | Four | 5 |
| | 11e | | More than four | 6 |

The threshold that determines whether two PACs are considered to belong to the same author oeuvre depends on the number of PACs belonging to an author name block. The larger this number, the higher the threshold. If there are many PACs that belong to the same author name block, there is a relatively high risk of incorrectly assigning two PACs to the same author oeuvre. To reduce this risk, a higher threshold is used. See Table 2 for used thresholds.

*Table 2: Relation between the number of PACs belonging to an author name block and the threshold used to determine whether two PACs are considered to belong to the same author oeuvre*

| Number of PACs in author name block | Threshold |
|---|---|
| 1 | - |
| 2-500 | 11 |
| 501-1500 | 13 |
| 1501-7000 | 17 |
| 7001-22500 | 21 |
| ≥ 22501 | 90 |

Figure 2 provides an illustration of the rule-based scoring and oeuvre identification phase. There are six PACs. The figure shows the result of applying the scoring rules combined with a threshold of 10 points. The score of PACs 1 and 2 equals 13 points. This is above the threshold value and, therefore, there is strong direct evidence that PACs 1 and 2 belong to the same author oeuvre. The same applies to PACs 2 and 3, PACs 3 and 4, and PACs 5 and 6. For other pairs of PACs, there is insufficient direct evidence to conclude that the PACs belong to the same author oeuvre. This is for instance the case for PACs 3 and 5. The scoring rules yield a score of 3 points for these PACs, which is below the threshold of 10 points. In the end, two candidate author oeuvres are obtained, one consisting of PACs 1, 2, 3, and 4 and the other one consisting of PACs 5 and 6. PACs 1, 2, 3, and 4 are assigned to same candidate author oeuvre because they belong to the same connected component in the network shown in Figure 2. Indirectly, there is strong evidence that PACs 1, 2, 3, and 4 all belong to the same author oeuvre.

*Figure 2: Network of PACs (threshold value of 10 points)*

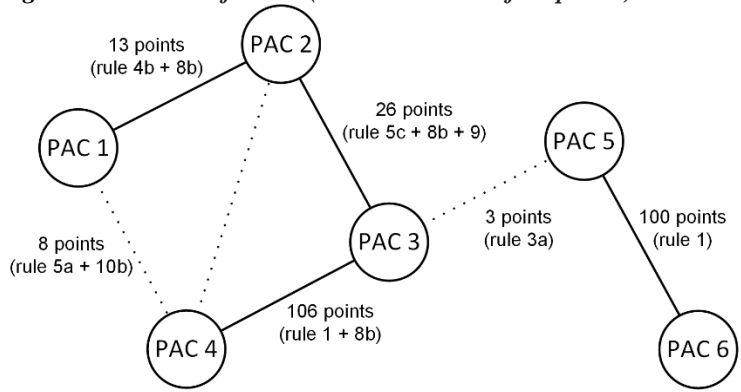

*Post-processing phase*

In the previous phase, candidate author oeuvres were identified separately for each author name block. In some cases, candidate author oeuvres obtained for different author name blocks need to be merged. This is for instance the case for an author that uses the name "Bernelli-Zazzera, Franco" in some of his publications and the name "Bernelli, Franco" in others. In the post-processing phase, candidate author oeuvres are merged if they share the same e-mail address. In this way, the final author oeuvres are obtained. In the remainder of this paper, we refer to the final author oeuvres as clusters.

When the final author oeuvres have been obtained, meta-data is generated for each of the associated clusters. Table 3 lists the fields included in the meta-data.

The CvE method values precision over recall: if there is not enough proof for joining publications together, the method will segregate them into separate clusters. As a consequence, the oeuvre of an author may be split over multiple clusters. The evaluation of the method carried out by Caron and Van Eck (2014) based on two datasets of Dutch researchers shows on average a precision of 95% and a recall of 90%, with the errors increasing for more common author names.

*Table 3: Fields list of the CvE clusters*

| Field | Description |
| --- | --- |
| cluster_id | Cluster identifier |
| n_pubs | Number of publications in the cluster |
| first_year | Cluster's earliest publication year |
| last_year | Cluster's latest publication year |
| full_name | Most common full name in cluster |
| first_name | Most common first name in cluster |
| email | Most common email address in cluster |
| address_organization | Most common organization in cluster |
| address_city | Most common city in cluster |
| address_country | Most common country in cluster |
| alternative_full_name | Second most common full name in cluster |
| alternative_first_name | Second most common first name in cluster |
| alternative_email | Second most common email address in cluster |
| alternative_address_organization | Second most common organization in cluster |
| alternative_address_city | Second most common city in cluster |
| alternative_address_country | Second most common country in cluster. |

## 2.2 The DGA heuristic approach to author name disambiguation

The DGA approach is based on the integration of a bibliometric database with an external database (D'Angelo, Giuffrida, & Abramo, 2011). The bibliometric database is the Italian National Citation Report, containing all WoS articles by those authors who indicated Italy as country of their affiliation, while the external source for data is the MIUR database described in Section 3.1. Figure 3 depicts the multi-stage process of the DGA approach, consisting mapping generation as the first step and filtering as the second.

*Figure 3: Flowchart of the DGA approach*

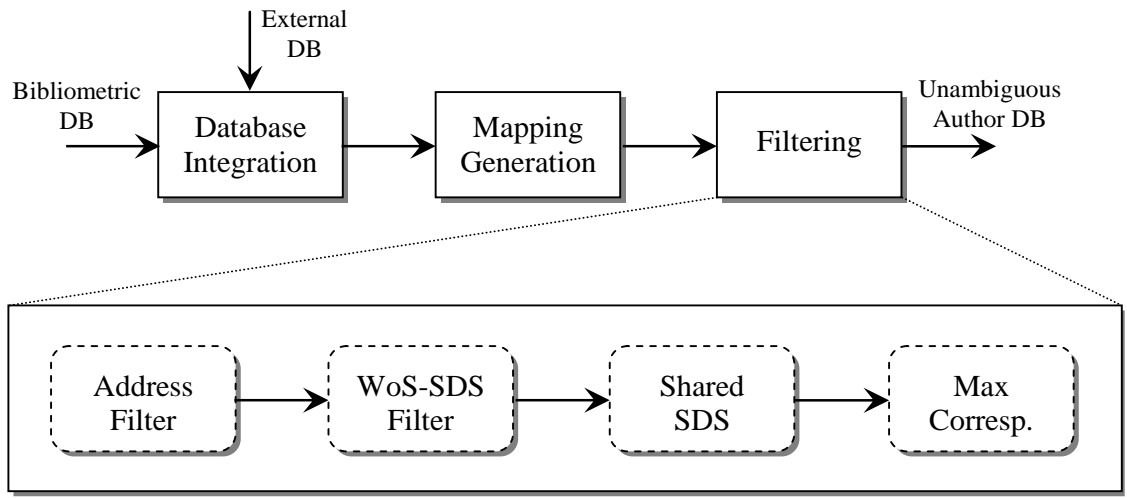

The objective of the first phase is to generate a mapping of the "authors" present in the bibliometric database and the "identities" indexed in the external database, through strategies of aggressive matching of last name and first name initials. The output is a series of author-identity pairs containing, for every author in the bibliometric database, different possible identities indexed in the external database. Note that the identity of each author is defined on an annual basis, since the external database indexes' personal information at the close of each year, without any correlation among identities that may pertain to different years.

This first phase generates both correct pairs but also a number of false positives because of all the possible cases of homonyms that the algorithm needs to eliminate through a step-by-step process, gradually filtering out undesired pairs. The filters employed follow data-driven heuristics. The first one is the "address filter", which eliminates all the author-identity pairs in which the author's affiliation (extracted from the "address" field of the bibliometric record) is incompatible with the identity's affiliation (the university identified for the researcher as listed in the external database). The effectiveness of the filter depends on the criteria employed for matching between the two fields, which are typically indicated in much different formats. The proposed algorithm employs rule-based criteria for matching based on a controlled vocabulary. From all the author-identity pairs that remain after the previous filter, the "WOS-SDS filter" eliminates all those in which the WoS subject category of the article published by the author is not compatible with the field associated to the identity in the external database. The idea is that an author who publishes an article in a certain subject category cannot possibly be associated with an identity that works in a completely different field. Again, in this case, the effectiveness of the filter depends on the criteria for matching the two classifications. The proposed algorithm carries out the matching in a deterministic fashion based on a purpose-prepared WoS-SDS mapping set. The filter is conceived to capture and remove obvious cases of homonyms revealed by evident incompatibility of the disciplinary categories, so as to minimize the production of false negatives.

Subsequently, more aggressive criteria for filtering are applied to the authors mapped with multiple identities that have survived the preceding filters. These obviously contain at least one false positive, which subsequent filters are designed to eliminate. The "shared SDS" filter chooses the identity corresponding to the SDS of a co-author that is already disambiguated. The idea is that a publication is more likely the result of collaboration between co-authors with the same SDS.

The "maximum correspondence filter" is finally used to process all the remaining authors mapped with multiple identities and, thus, address all the remaining cases of unresolved homonyms. In this case, the filter chooses the pair for which the identity's SDS has maximum "correspondence" to the subject category of the article. The correspondence of an SDS to a particular subject category is defined (on the basis of a seed set) as the number of identities belonging to that SDS that result as authors of articles falling in the subject category. The algorithm uses a seed set constructed in an automatic fashion based on the authors of all the pairs already accepted as correct by the algorithm.

In the original paper (D'Angelo, Giuffrida, & Abramo, 2011), the DGA approach was tested on: 1) a sample of 372 Italian publications, resulting in a precision of 95.6% and a recall of 93.8%; and 2) the institutional publication list of professors affiliated to the University of Milan, resulting in a precision of 96.4% and a recall of 94.3%.

## 3. Methodology

We propose to use the CvE method to first extract relevant publication clusters and, then, in a subsequent step, filter and merge the extracted publication clusters by means of a reference institutional database, specifically the one used in the DGA method. In the following subsections, we will illustrate the dataset used in the analysis and the adopted procedure.

## 3.1 Dataset

We carried out an empirical analysis on a sample of Italian professors. The data source is the database maintained by the Ministry of Education, Universities and Research (MIUR)[2], indexing the full name, academic rank, research field and institutional affiliation of all professors at Italian universities, at the close of each year. Observed at 31 December 2016, there were 52,861 full, associate, and assistant professors working at Italian universities. Each professor is classified in one and only one of the 370 research fields referred to as "scientific disciplinary sectors" (SDSs).[3] The SDSs are grouped into 14 disciplines known as "university disciplinary areas" (UDAs). To ensure the robustness of the bibliometric approach, our reference population is limited to the 36,211 professors in the science sectors in which the research output is likely to be extensively indexed in the WoS. From this population, 615 professors (145 full, 228 associate, 242 assistant) from 71 different Italian universities have been randomly selected. This sample assures a projection of the precision and recall values on the whole population, with a margin of error of no more than ±2%, at a 95% confidence level. Table 4 shows the distribution by disciplinary area of all Italian professors and professors included in the random sample.

*Table 4: Distribution by disciplinary area of all Italian professors and the professors included in the random sample*

| Area | All Italian professors | Random sample |
|---|---|---|
| Mathematics and computer science | 2,918 (8.1%) | 46 (7.5%) |
| Physics | 2,062 (5.7%) | 23 (3.7%) |
| Chemistry | 2,714 (7.5%) | 50 (8.1%) |
| Earth sciences | 974 (2.7%) | 18 (2.9%) |
| Biology | 4,471 (12.3%) | 88 (14.3%) |
| Medicine | 8,746 (24.2%) | 147 (23.9%) |
| Agricultural and veterinary sciences | 2,880 (8.0%) | 45 (7.3%) |
| Civil engineering and architecture | 1,520 (4.2%) | 26 (4.2%) |
| Industrial and information engineering | 5,170 (14.3%) | 91 (14.8%) |
| Pedagogy and psychology | 1,350 (3.7%) | 22 (3.6%) |
| Economics and statistics | 3,406 (9.4%) | 59 (9.6%) |
| Total | 36,211 | 615 |

To get an idea of the complexity of the disambiguation of author names in the context in question, in Table 5, we show the frequencies of the potential cases of homonymy related to the 615 professors in our sample with respect to the whole Italian academic population. Only 71% of the professors (438 in total) do not have potential homonyms among their colleagues in the national academic system. Another 87 show at least one homonym, 31 two, and 17 three. For 23 out of the 615 professors in the sample, we registered at least 6 homonyms. In this regard, Table 6 reports the 10 most complex cases: "Rossi, Fausto" holds the record with a last name and first initial combination ("Rossi, F") that is shared with 40 other professors at Italian universities.

---

[2] http://cercauniversita.cineca.it/php5/docenti/cerca.php, last accessed 20/09/2019.
[3] The complete list is accessible at attiministeriali.miur.it/userfiles/115.htm, last accessed 20/09/2019.

*Table 5: Distribution of the number of homonyms in the Italian academic system for the 615 professors included in the dataset*

| No. of homonyms in the Italian academic system | Frequency |
|---|---|
| 0 | 438 |
| 1 | 87 |
| 2 | 31 |
| 3 | 17 |
| 4 | 13 |
| 5 | 6 |
| 6 or more | 23 |
| Total | 615 |

*Table 6: Number of homonyms in the Italian academic system for the professors in the dataset with the most common names*

| Name of the sampled professor | Corresponding author name | No. of homonyms in the Italian academic system |
|---|---|---|
| ROSSI, Fausto | ROSSI, F | 40 |
| RUSSO, Antonio | RUSSO, A | 35 |
| MARTINI, Marco | MARTINI, M | 16 |
| ROMANO, Mario | ROMANO, M | 16 |
| RICCI, Francesco | RICCI, F | 13 |
| FERRARA, Maria | FERRARA, M | 12 |
| GATTI, Marco | GATTI, M | 11 |
| LEONE, Antonio | LEONE, A | 11 |
| MARINI, Amedeo | MARINI, A | 11 |
| ROMANO, Severino | ROMANO, S | 10 |

### 3.2 Procedure

For each of the 615 professors in the sample, the 2010–2016 WoS publication portfolio was collected through the following methods:
- The extraction of publication clusters based on the CvE author name disambiguation process, as described in Section 3.2; and
- The filtering of extracted clusters based on information retrieved from the external MIUR database. This filtering is inspired by the DGA method described in Section 3.3.[4]

Regarding the first step, cluster extraction was achieved through matching of all possible combinations of last name and first name initials. For example, for "BERNELLI ZAZZERA, Franco" we checked "bernelli, f%", "zazzera, f%", "bernellizazzera, f%", "bernelli zazzera, f%", and "bernelli-zazzera, f%",[5] and extracted in this way the eight clusters shown in Table 7.

---

[4] Note that this filtering stage differs from that in the original DGA method: in the original DGA method, author-identity pairs are filtered; in the proposed approach, complete clusters are filtered.
[5] The percent sign (%) wildcard allows to retrieve any name starting with the text preceding the sign.

*Table 7: CvE clusters extracted by querying for all possible combinations of last name and first name initials of "BERNELLI ZAZZERA, Franco"*

| cluster_id | n_pubs | first_year | last_year | full_name | first_name |
|---|---|---|---|---|---|
| 7791208 | 1 | 2003 | 2003 | bernelli, f | |
| 7791209 | 35 | 1989 | 2016 | bernelli-zazzera, f | franco |
| 41033608 | 1 | 2000 | 2000 | bernelli-zazzera, f | |
| 41033609 | 1 | 2002 | 2002 | bernelli-zazzera, f | |
| 41033610 | 1 | 2005 | 2005 | bernelli-zazzera, f | |
| 22689350 | 1 | 2008 | 2008 | zazzera, f | francesca |
| 22689348 | 2 | 2014 | 2015 | zazzera, fb | franco bernelli |
| 22689349 | 1 | 2007 | 2007 | zazzera, fb | f. bernelli |

In addition to the fields shown in Table 7, every single cluster is fully described in terms of its most common author data, for a total of the 16 fields shown in Table 3. In short, each cluster contains a certain number of publications (n_pubs) attributed to a certain author within a certain time window (first_year; last_year). Based on this information, we can remove the clusters characterized by a time window with an empty intersection with the 2010–2016 period. Looking at Table 7, this means that for "BERNELLI ZAZZERA, Franco" we can further consider only those clusters with cluster_id 7791209 and 22689348.

Overall 9,069 clusters were retrieved, related to 603 professors, indicating that for 12 (2%) professors (out of in total 615) in the sample, no clusters were found. For 179 (29%) professors, the queries retrieved one single cluster. For the remaining 424 (69%) sampled professors, the queries returned more than one cluster, shown in Figure 4, and, specifically more than 10 clusters for 19% of the professors, more than 50 clusters for 5% of the professors. Finally, 51 clusters were assigned to two distinct homonyms:

- MANCINI Francesco, professor of Clinical Psychology at the "Guglielmo Marconi" University in Rome; and
- MANCINI Francesco Paolo, professor in Biochemistry at the University of Sannio in Benevento.

*Figure 4: Relative frequencies of number of CvE clusters retrieved for 615 professors in the sample*

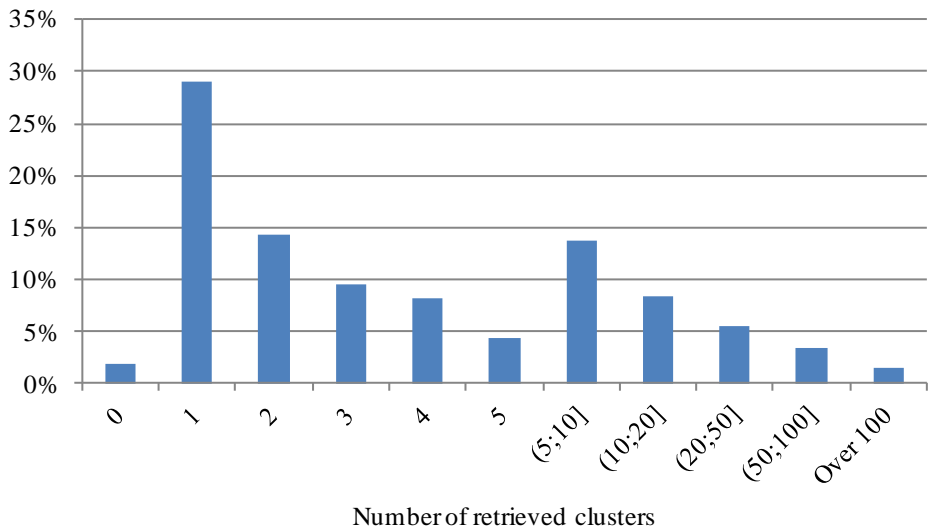

The 9,069 clusters retrieved as described above were filtered according to three distinct scenarios.

*Scenario 1:* We removed clusters for which the most occurring country (address_country) and the second most occurring country (alternative_address_country) were different from "Italy". This avoids false positives due to foreign homonyms, but causes false negatives related to publications in which the author appears only with a foreign affiliation.[6] To maximize recall, we included clusters without address_country information. We also removed the clusters where the complete first name of the author (where available) was "incompatible" with that of the considered professor (e.g., "Franco" vs "Federico").

*Scenario 2:* In addition to Scenario 1, we added a filter based on the city (address_city or alternative_address_city) of the university to which the subject in the sample was affiliated on 31/12/2016. To maximize recall, we included clusters without address_city information.

*Scenario 3:* We performed a "manual" validation of all retrieved clusters, without any kind of automatic filtering, but using the information provided by the MIUR database about the career of each sampled subject[7].

There is an evident trade-off between the cost/effort required to implement the filtering process and the resulting level of accuracy of these three scenarios. In fact, Scenario 1 is the easiest and cheapest to implement, but is characterized by a low precision due to the low capability to filter false positives caused by national homonyms. In contrast, Scenario 3 should guarantee maximum accuracy, since any possible false is caused only by human error. However, the manual validation is extremely expensive and, above all, unfeasible on large scale datasets. Finally, Scenario 2 should guarantee intermediate levels in terms of both cost and precision/recall of the retrieved portfolios. In particular, it requires only the knowledge of the city where the organization to which the author belongs is located. Of course, this kind of filtering can generate false negatives in the case of subjects with a high "mobility" in the considered publication period. However, compared to Scenario 1, it should ensure a higher level of precision, thanks to a higher capability to filter false positives national homonyms.

## 4. Results and analysis

As shown in the last row of Table 8, the filtering process drastically reduces the initial 9,069 clusters to 2,057 clusters in Scenario 1, 1,276 clusters in Scenario 2, and 1,256 clusters in Scenario 3. As indicated above, the initial number of clusters assigned to a professor varies largely. 179 professors are assigned to only one cluster, while 110 professors are assigned to 12 or more clusters. The filtering stages applied in the three scenarios, substantially change the distribution of professors over the number of assigned clusters. In Scenario 1, 305 professors are assigned to a unique cluster and 26 professors are assigned to 12 or more clusters. One professor is assigned to no more than 136 clusters. Scenario 3 seems to be the most accurate with 383 professors assigned to a unique cluster. Also in this case, however, the multiple cluster assignments are numerous, affecting one third of professors in the sample, with ten having 12 or more clusters and one, even 109 clusters. To some extent, these results offer a quantitative measure of what the authors of the CvE approach mean when they say, "if there is not enough proof for joining publications together, they will be put in separate clusters. As a consequence, the oeuvre of an author may be split over multiple clusters" (Caron & van Eck, 2014). Finally, Scenario 2 seems "intermediate" between the two, but registers 42 professors without any clusters assigned.

---

[6] In fact, these publications are also ignored by the DGA algorithm, which is applied only to articles indexed in the Italian National Citation Report

[7] The reader may wonder why a "manual validation" is performed in an approach proposed for "large scale" author name disambiguation. As we will see better below, this scenario is presented only to understand the trade-off between costs and benefits of this scenario and the less costly alternative scenarios in which no manual validation is involved.

*Table 8: Distribution of professors over the number of assigned clusters, for each scenario*

| No. of clusters assigned | No. of professors | | | |
|---|---|---|---|---|
| | No filtering | Scenario 1 | Scenario 2 | Scenario 3 |
| 1 | 179 | 305 | 376 | 383 |
| 2 | 88 | 137 | 123 | 120 |
| 3 | 59 | 46 | 25 | 32 |
| 4 | 50 | 24 | 16 | 17 |
| 5 | 27 | 14 | 10 | 9 |
| 6 | 19 | 11 | 5 | 5 |
| 7 | 20 | 10 | 3 | 0 |
| 8 | 11 | 1 | 1 | 0 |
| 9 | 23 | 4 | 1 | 1 |
| 10 | 11 | 5 | 1 | 2 |
| 11 | 5 | 5 | 0 | 2 |
| 12 ore more | 110 | 26 | 12 | 10 |
| Total | 603 | 588 | 573 | 581 |
| Max | 2341 | 136 | 105 | 109 |
| Without clusters | 12 | 27 | 42 | 34 |
| Total distinct clusters | 9069 | 2057 | 1276 | 1256 |

To check for the accuracy of the census of the publication portfolio of the 615 sampled professors, we used a reference dataset containing disambiguated publications authored in the observed period (2010–2016) by these professors. Having started from a randomly extracted sample and not from an existing standard bibliometric dataset, we needed to build the "reference" dataset with an ad hoc procedure. Aiming at minimizing (and possibly having zero) possible false positives and negatives with respect to the real overall scientific production of each of the 615 professors, we proceeded in generating redundancy by combining the results of the application of several approaches. More specifically, our reference dataset has been obtained by manually checking and merging the following:
- Authorships related to the 2,084 distinct clusters obtained by the three filtering scenarios described above;
- Authorships obtained by applying the DGA algorithm to documents indexed in the Italian National Citation Report; and
- Authorships identified by querying the WoS using the ORCID of each of the sampled professors[8].

The reference dataset contains 11,672 authorships, related to 11,206 publications authored by 577 (out of 615) professors in the sample.[9] The difference between the number of authorships and the number of publications is due to 464 publications co-authored by two distinct sampled professors and one by three.

---
[8] All Italian university research staff hold an ORCID identifier, following the IRIDE project launched in 2014 by the MIUR.
[9] We have excluded documents published in a year in which the relevant author was not a tenured professor in the Italian academic system.

Table 9 shows the precision, recall, and F-measure obtained by:
- Filtering (according to the three scenarios described above) the clusters obtained through the CvE disambiguation approach (columns 2–4);
- Applying the DGA algorithm as a baseline (column 5); and
- Applying two other baseline methods (columns 6 and 7), tagged as Baseline 1, where name instances are clustered based on their last name and first name initials, and Baseline 2, where name instances are clustered based on their last name and full first name (Backes, 2018; Kim & Kim, 2019).[10]

We want here to remind that

$$\text{Precision} = \frac{(\text{Relevant authorships} \cap \text{Retrieved authorships})}{\text{Retrieved authorships}}$$

$$\text{Recall} = \frac{(\text{Relevant authorships} \cap \text{Retrieved authorships})}{\text{Relevant authorships}}$$

$$\text{F-measure} = 2\frac{\text{Precision} \cdot \text{Recall}}{\text{Precision} + \text{Recall}}$$

with

$$\text{Relevant authorships} = \text{Retrieved authorships} + \text{false negatives} - \text{false positives}$$

As expected, Scenario 3 is actually the most accurate, with a precision of 96.9% and a recall of 97.4%. Scenario 1 shows a similar recall (97.6%) but a much worse precision (76.6%) due to the large number of false positives. The performance of Scenario 2 seems very interesting. Considering the limited effort needed to implement such a filtering strategy, we obtain a very high F-measure (96.1), more than two points higher than that obtained through the DGA baseline method (93.9%). Compared to the other two baseline methods, it can be seen that the performance of Scenario 2 is similar to that of Baseline 1 in terms of recall (96.0% vs 96.1%), but it is clearly better in terms of precision than both Baseline 1 (96.1% vs 44.2%) and Baseline 2 (96.1% vs 89.2%).

*Table 9: Performance of the contrasted approaches, measured on the sampled professors*

|  | CvE Scenario 1 | CvE Scenario 2 | CvE Scenario 3 | DGA | Baseline1 | Baseline2 |
|---|---|---|---|---|---|---|
| Retrieved authorships | 14875 | 11659 | 11743 | 11725 | 25351 | 11730 |
| False positives | 3485 | 450 | 369 | 736 | 14138 | 1272 |
| False negatives | 282 | 463 | 298 | 683 | 459 | 1214 |
| Precision | 76.6% | 96.1% | 96.9% | 93.7% | 44.2% | 89.2% |
| Recall | 97.6% | 96.0% | 97.4% | 94.1% | 96.1% | 89.6% |
| F-measure | 85.8% | 96.1% | 97.2% | 93.9% | 60.6% | 89.4% |

However, these aggregate results do not tell us if false positives and negatives are concentrated or spread over the sampled subjects. For this reason, Figure 5 provides histograms for the three scenarios applied to filter the clusters obtained with the CvE approach. These histograms show the frequency distribution for different ranges of the F-measure obtained for individual professors in the dataset. The percentage of the subjects with no errors (an F-measure of 100%) varies from a minimum of 61.8% in Scenario 1 to a maximum of 77.3% in Scenario 3. For Scenario 1, 54 (9.2%) professors have an F-measure less than 60%, of which 20 have an F-measure less than 10%. In Scenario 2, the maximum accuracy (an F-measure of 100%) is registered for 74.3% of the professors. Here, 26 (4.5%) professors have an F-measure less than 60%, of which 18 show an F-measure less than 10%.

---

[10] Baseline 1 is a simple method often performed by scholars in practice. Given the high share of potential homonyms (29% as shown in Table 5), we expect a low level of precision when applying such method. Baseline 2 should solve most homonym cases but could lead to a low level of recall due to an increasing number of false negatives.

*Figure 5: F-measure of the CvE approach at the individual level, in the three considered scenarios*

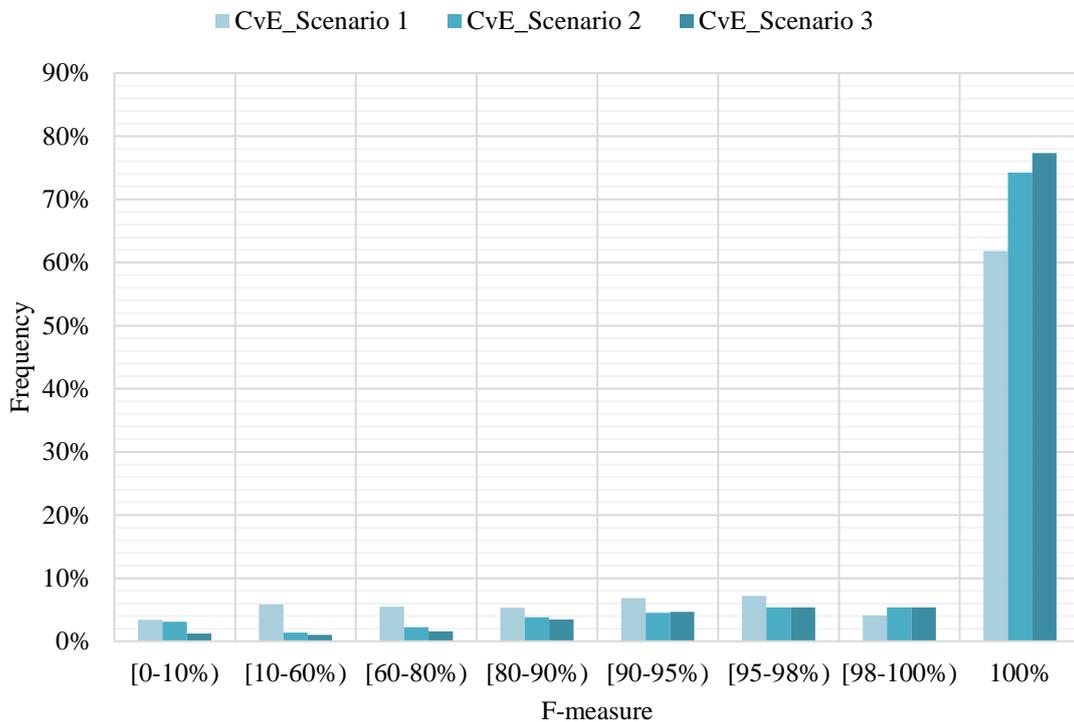

Comparing CvE Scenario 2 to DGA, Table 9 shows a difference of 2.4% for precision and 1.9% for recall, in favor of the former. Regarding precision, the analysis at the individual level reveals a substantial similar performance for the two approaches (Figure 6). Focusing on the left tail of the distribution, CvE Scenario 2 shows a somewhat higher percentage of cases with low precision levels, i.e. lesser than 60%. This can be due to the low capability of this approach to filter false positives due to homonyms working in the same city. These cases are better managed by the DGA approach, which applies additional filters based on the correspondence of the subject category of the publication to the SDS of the subject.

The distribution of recall obtained at the individual level shows however the clear superiority of the CvE Scenario 2 approach (Figure 7), with a 100% recall registered for 84.1% of the sampled subjects, against 65.6% for the DGA approach, which generates at least one false negative in almost 35% of the sampled subjects. An in-depth analysis of the possible causes of such false negatives reveals that:

- In 21.6% of the cases, the subject in the byline was not identified, i.e. no author-identity pairs were generated in the first mapping stage,
- In 47.6% of the cases, the correct pair was wrongly eliminated by the address filter, since no bibliometric address had been matched to the academic affiliation of the subject, and
- In 30.8% of the cases, the false negative was originated by the application of the WoS-SDS filter or other filters based on the correspondence between the subject category of the publication and the SDS of the author.

As for the first two causes, the CvE approach seems to be more robust because it does not apply a binary logic on a single bibliometric metadata element but a continuous score based on a combination of different bibliometric metadata elements. As for the third cause, it is evident that this kind of filter reduces false positives but, at the same time, generates false negatives when authors occasionally vary their scientific activity by publishing on topics not included in the core of their reference field.

*Figure 6: Precision of the CvE Scenario 2 and DGA approaches at the individual level*

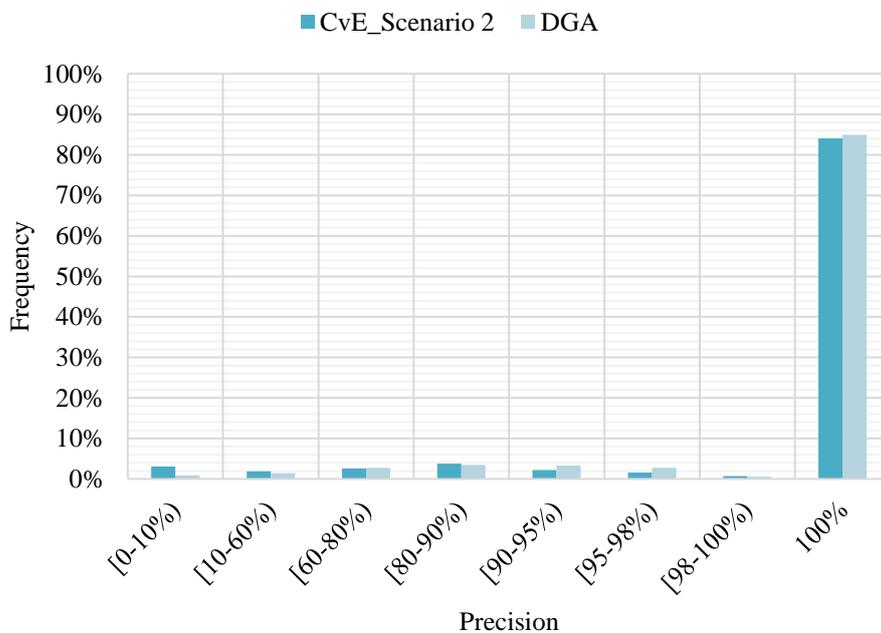

*Figure 7: Recall of the CvE Scenario 2 and DGA approaches at the individual level*

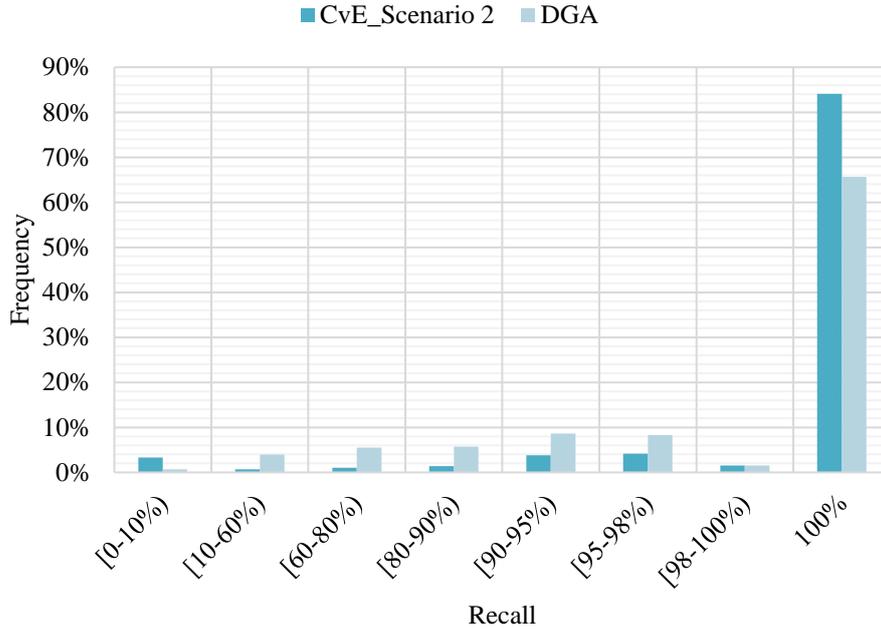

## 5. Conclusions

The quality of the bibliometric dataset on which a research evaluation exercise is based is crucial. In large-scale assessments, the different data collection options have to be evaluated in terms of the trade-off between accuracy and costs, including the opportunity costs when the surveyed subjects are asked to collect and select the research outputs to be evaluated. Actually, indirect costs in general are estimated to be much higher than direct costs and can be minimized (if not completely saved) only if the evaluator proceeds by autonomously selecting the publications produced by the subjects from the relevant bibliometric databases. This option offers rapid and economical implementation but is also very challenging if the evaluator wants to rely on a very accurate census of the scientific portfolio of the assessed units, given the technical complexity of disambiguating the true identity of authors in the byline of publications. Both supervised and unsupervised methods proposed in the literature for this purpose show critical issues and generally favor precision over recall. In this paper, we have proposed a new approach that relies on an external source of information for selecting and validating clusters of publications identified using the CvE unsupervised author name disambiguation method.

We applied the proposed approach to a sample of 615 Italian scholars and measured the accuracy of the census of their publication portfolio to verify the generalizability of a disambiguation procedure relying on an external source containing few essential data on the subjects to be evaluated.

The obtained results are particularly encouraging:
- By knowing the complete first name of the subject and their exact affiliation city, we obtained a census with an overall F-Measure equal to 96.1% (96.1% for precision; 96.0% for recall), 2% higher than that recorded by applying the DGA baseline approach.
- The 4% error is not evenly distributed among the observed subjects: for 74.3% of

them, the census is perfectly accurate (an F-measure of 100%). Critical cases (meaning those with an F-measure less than 60%) amount to 4.5% out of the total.
- The error distribution also seems to be much more favorable than the one resulting from the DGA baseline approach, especially in terms of recall.

The measured performances are not independent of the considered time window. By increasing the time window, the likelihood of the "mobility" for individual subjects will increase and the recall reduce due to false negatives generated by the application of a "static" city filter. The considered time window of 7 years is fully compatible though with national research evaluation exercises and many other relevant evaluative frameworks. Therefore, we dare to conclude that the approach proposed in this study could be used as a starting point for those in charge to carry out large scale census of publication portfolios (research managers, policy makers and evaluators in general) for bibliometric research evaluation purposes, especially at the individual level.

The external source of information, albeit crucial for the applicability of our approach, is not a particularly critical resource. National and international research systems are typically composed of communities that can be easily identified, and gathering data to build a comprehensive external database should not require significant human efforts, especially considering that it should contain only full personal names and affiliation cities of the subjects to be assessed. Of course, it should be noted that the approach proposed in this paper has been evaluated on researchers affiliated to Italian universities. Name ambiguity issues vary across country and ethnicity. As reported in several studies, East Asian researcher names have been found to be challenging due to many homonym cases (Strotmann, & Zhao, 2012). If tested on different types of ethnic names, the reported performance of the proposed approach might be different. With our proposal, we hope to arouse the curiosity of scholars who are interested in reproducing such an analysis in other national contexts.

Finally, we would like to emphasize that research evaluations at the individual researcher level are difficult and delicate to carry out and need to be performed with care: errors are possible and can affect career, funding, or similar critical decisions. Nonetheless, individual evaluations are carried out, continuously, every day, very often with heavy manual work to collect publication data. In this paper, we tried to propose a semi-automated approach and supplied a quantitative measure of the associated errors. In the end, the evaluator has to judge whether these errors are within acceptable limits or not, given the consequence of the study and the evident trade-off between the accuracy of data and the costs that are needed to achieve it.